\documentclass[prb,twocolumn,showpacs,floatfix,groupeaddress]{revtex4}
\usepackage{amsmath,amsfonts}
\usepackage{graphicx}
\usepackage{color}
\usepackage{enumerate}

\newcommand{\e}{\mathrm{e}}

\newcommand{\up}{\uparrow}
\newcommand{\dw}{\downarrow}

\begin{document}


\title{Non-Kondo many-body physics in a Majorana-based Kondo type system}

\author{Ian J. van Beek}

\author{Bernd Braunecker}
\affiliation{SUPA, School of Physics and Astronomy,
	University of St Andrews, North Haugh, St Andrews KY16 9SS, UK}

\date{\today}




\begin{abstract}
We carry out a theoretical analysis of a prototypical Majorana system, which demonstrates the existence of a Majorana-mediated many-body state and an associated intermediate low-energy fixed point. Starting from two Majorana bound states, hosted by a Coulomb-blockaded topological superconductor and each coupled to a separate lead, we derive an effective low-energy Hamiltonian, which displays a Kondo-like character. However, in contrast to the Kondo model which tends to a strong- or weak-coupling limit under renormalization, we show that this effective Hamiltonian scales to an intermediate fixed point, whose existence is contingent upon teleportation via the Majorana modes.  We conclude by determining experimental signatures of this fixed point, as well as the exotic many-body state associated with it.

\end{abstract}


\maketitle


Majorana zero modes have attracted significant interest in recent years, \cite{alicea2012,leijnse2012,beenakker2013,aasen2015}
due in no small part to their potential for realizing topologically protected
quantum computing architectures. \cite{kitaev2003,nayak2008}
A variety of systems have been proposed, and in part experimentally implemented, to host
Majorana modes, including superconductor contacted topological insulators, \cite{fu2008}
semiconductor nanowires with strong spin-orbit interaction,
\cite{sato2009,oreg2010,lutchyn2010,mourik2012,deng2012,das2012,rokhinson2012,finck2013,churchill2013}
magnetic adatom chains on superconductors
\cite{nadj-perge2013,braunecker2013,klinovaja2013,vazifeh2013,pientka2013,poyhonen2014,rontynen2014,kim2014,heimes2014,nadj-perge2014,brydon2015,poyhonen2015,pawlak2015,ruby2015,braunecker2015,schecter2016}
and coupled Josephson junction arrays. \cite{vanHeck2011,vanHeck2012,hassler2012}

\begin{figure}
	\centering
	\includegraphics[width=\columnwidth]{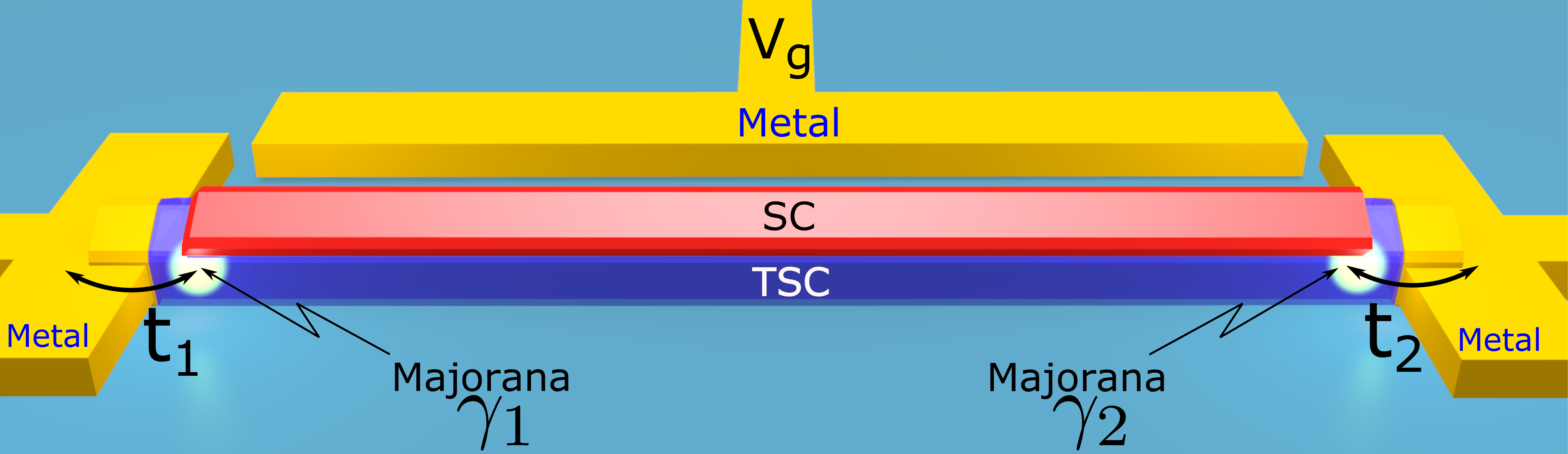}
	\caption{\label{setup}
		A Majorana system with a floating superconductor. An $s$-wave superconductor (red) is grown on a nanowire
		with strong spin-orbit coupling (blue). For sufficiently large applied magnetic field and appropriate
		chemical potential the nanowire becomes a topological superconductor (TSC) with Majorana bound states $\gamma_{1,2}$
		at each end. A gate at voltage $V_g$ is used to tune the ground state occupation number, which is dictated
		by the capacitive charging energy $E_C$. The nanowire couples to leads at either end with tunnelling
		coefficients $t_1$ and $t_2$ between lead electrons and $\gamma_{1,2}$.
	}
\end{figure}

In this paper we assume the existence of such a Majorana system, for instance, in the nanowire setup depicted
in Fig. \ref{setup}, in which Majorana modes appear at the wire ends as a result of spin-orbit interaction in the wire, an applied magnetic field and superconductivity induced through contact to an $s$-wave superconductor. \cite{sato2009,oreg2010,lutchyn2010}
Furthermore, we consider a floating superconductor so that there is a charging energy $E_C$ associated with the tunnelling
of electrons to and from the nanowire. Several studies have been performed on the low energy behaviour of such a system,
predicting distinctive non-local transport and Coulomb-blockade phenomena.
\cite{fu2010,zazunov2011,hutzen2012,zazunov2012,wang2013,ulrich2015,sau2015,plugge2015,plugge2016}
The latter of these appears to have been confirmed in a recent experiment. \cite{albrecht2016}
A coupling of several such Majorana wires through a common floating superconductor gives rise to the
topological Kondo effect.
\cite{beri2012,altland2013,beri2013,galpin2014,zazunov2014,altland2014a,altland2014b,kashuba2015,buccheri2015} This is a result of the existence of Majorana modes in combination
with constrained fluctuations due to a charging energy $E_C$. A Majorana mode may also be coupled to a quantum dot to explore the competition between Kondo and Majorana physics. \cite{lee2013,chirla2014,cheng2014,rt2015} However, these works do not fully explore the
potential of the Majorana modes as a novel interface between the topological superconductor
and its environment.

In this work, we show that such an exploration reveals fundamentally new physics
in which the Kondo and Majorana aspects combine and lead to a new type of many-body state. This marriage of traditionally distinct physics leads us to call the result the \emph{Kondorana} model.

We consider a single Majorana wire as shown in Fig. \ref{setup} and tune it
to degeneracy of two different charging states, such that tunnelling into the Majorana states
can make a transition between the degenerate states or lead to a high energy ($2E_C$)
excitation. By integrating out the latter, we
obtain an effective Kondo like low energy theory, in which the two degenerate charging states
take the role of the Kondo spin $S$. However, the situation differs from the Kondo
model in two essential ways. First, a Kondorana spin flip is induced by electron tunnelling
and not by an electron spin flip type process. Second, the effective $S^z$ interaction
couples not to the electron spin but to a pseudo-spin $s$ constructed from the electron operators for the
left and right leads. In addition to the regular $S^z s^z$ coupling, the teleportation
property of the Majorana states \cite{fu2010} leads to a further $S^z s^y$ coupling. Due to
the latter, the renormalization group flow for the interaction strength has a zero eigenvalue
and hence the fixed point of this Kondorana model is finite and does not lie at zero or
infinity as for the Kondo model. Nevertheless, this fixed point describes a many-body state extending
across the metallic leads, superconducting condensate, and Majoranas. It is important to note that the Kondorana fixed point is distinct from that found in the two channel Kondo Model, \cite{nozieres1980} despite superficial similarities arising from the invocation of Majorana modes to solve the latter scenario. \cite{coleman1995, emery1992} Finally,
we determine how the conductance of the nanowire scales with the ratio of tunnelling couplings and with the temperature,
and we suggest signatures of this state, that should be observable
with current experimental techniques.

After completion of this work we became aware of Ref. \onlinecite{bao2016} in which a similar setup was investigated in a time reversal invariant topological superconductor with two Majorana states at each end of the wire. Remarkably, instead of Kondorana physics, a two channel Kondo model is obtained in this system.    


\section{Model} 
\begin{figure}
	\centering
	\includegraphics[width=\columnwidth]{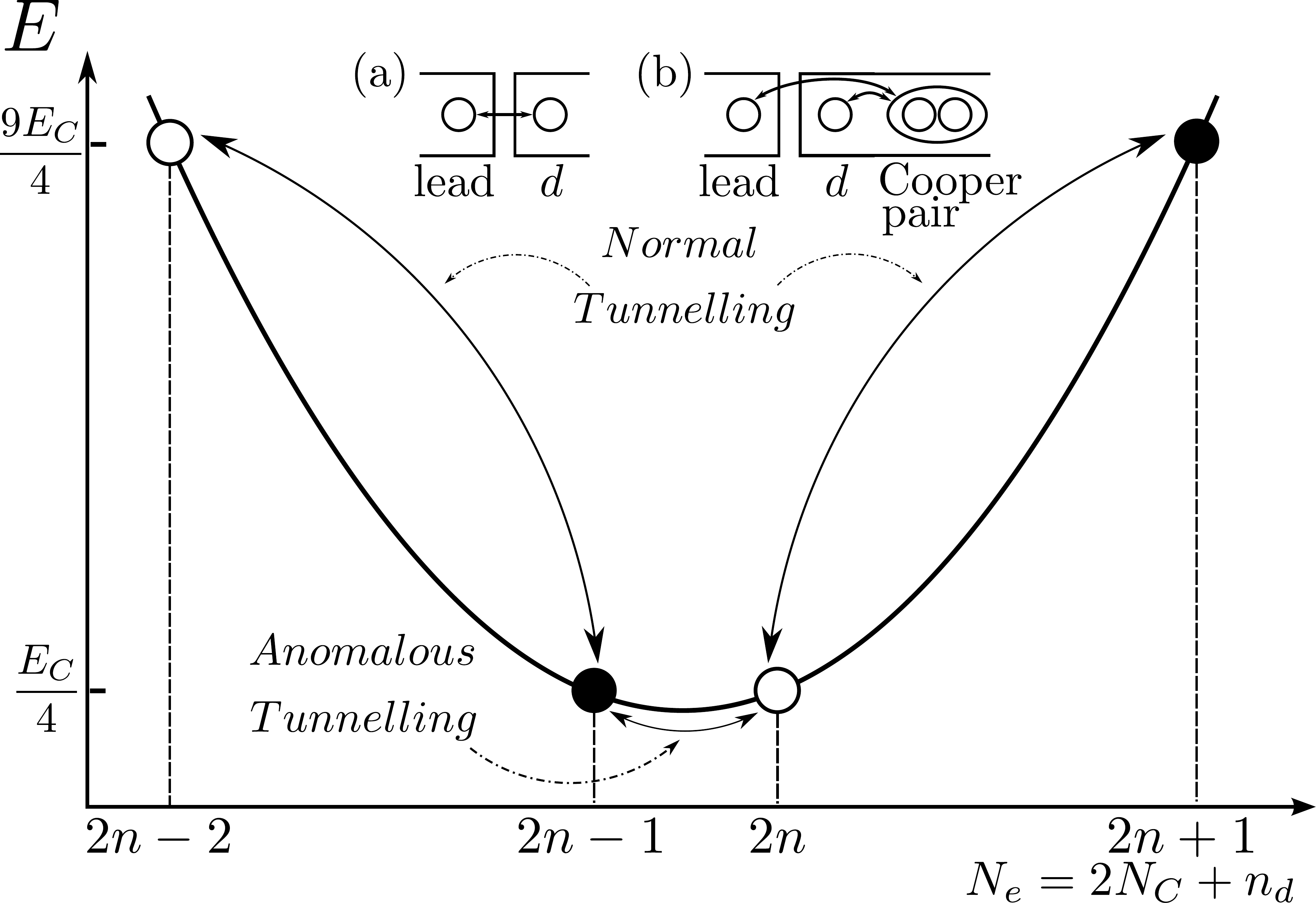}
	\caption{\label{elandscape}
		Charging energy against total number of electrons, $N_e$, in the nanowire for $n_g = 2n-\frac{1}{2} $.
		Filled (empty) circles represent states with $n_d=1$ ($n_d=0$). Both the ground and excited states are degenerate,
		with transitions between them via normal tunnelling (a)
		and anomalous tunnelling (b). Note that there is no process mediating
		transitions between the two excited states.
	}
\end{figure}

Our analysis is based on the Majorana Single Charge Transistor (MSCT), \cite{zazunov2011,hutzen2012}
which results from the usual Majorana setup of a quantum wire with strong spin-orbit interaction in a
magnetic field, but where the coupled superconductor is mesoscopic and floating, with a charging energy $E_C$,
where $E_C \ll \Delta_{TS}$,
with $\Delta_{TS}$ being the proximity induced gap of the topological superconductor. We furthermore
assume that $E_C$ is large compared with all other energy scales, notably the tunnel couplings $t_1$ and $t_2$ to
the leads, temperature $T$ and applied voltage bias $V$. This system is described by the Hamiltonian
$H = H_{el} + H_T + H_C$.
The leads are treated as non-interacting reservoirs,
$H_{el} = \sum_{j,k,\sigma} \epsilon_{jk} c_{jk\sigma}^\dagger c_{jk\sigma}$,
where $c_{jk\sigma}$ are electron operators for leads $j=1,2$, momenta $k$ and spins $\sigma=\up,\dw$, with the
dispersion $\epsilon_{jk}$. The coupling between the leads and the superconductor is restricted to
tunnelling into the Majorana states $\gamma_1$ and $\gamma_2$, and we explicitly exclude the possibility of exciting
quasiparticles. \cite{plugge2016}
The tunnelling Hamiltonian can then be written as \cite{hutzen2012}
$H_T = \sum_{k} ( t_1 c_{1k\dw}^\dagger \eta_1 + i t_2 c_{2k\up}^\dagger \eta_2 ) + \text{H.c.}$
We note that tunnelling through the Majoranas is spin polarized, \cite{sticlet2012,oreg2010}
for instance, with opposite spins for both Majoranas if the magnetic field is applied perpendicular to the spin-orbit
polarization direction, as written here. The spin polarization may also be non-antiparallel, if the magnetic field is tilted or if there exists a mixture of Rashba and Dresselhaus spin orbit coupling. For the purposes of this paper, it is only important that the coupling to the leads no longer has the spin degree of freedom. This allows us to effectively eliminate the spin index in the notations and we write $c_{1k}=c_{1k\dw}$ and $c_{2k}=c_{2k\up}$. Furthermore
$t_1$ and$t_2$ are the tunnelling amplitudes and
$\eta_{1,2} = d \pm \e^{-i\chi} d^\dagger$, with $d = (\gamma_1+i\gamma_2)/\sqrt{2}$. The form of the $\eta_{1}$ and $\eta_2$ operators takes into account that
tunnelling between Majorana and lead can occur over two channels: by removal of an electron from the fermionic state
$d$ obtained by the superposition of $\gamma_1$ and $\gamma_2$ (normal tunnelling), or by splitting a Cooper pair and transferring one electron to the lead
and the other electron to the $d$ state (anomalous tunnelling) as shown in Fig. \ref{elandscape}. The removal of a Cooper pair is expressed by
$\e^{-i\chi}$, where $\chi$ is the superconducting phase operator, which obeys $[N_C,\e^{-i\chi}]= -\e^{-i\chi}$
where $N_C$ is the Cooper pair number operator. We have deliberatley omitted Andreev tunnelling processes from our analysis for two reasons. First, their amplitude is proportional to  $t^2/\Delta$ and so is much smaller than the amplitude, $t$, relevant for the considered processes. Second, an Andreev process changes the number of charges on the nanowire by $\pm 2$, leaving the system in an excited state that needs further relaxation, and so the Andreev processes exist only at higher orders.  
Finally, the charging state of the Majorana wire is given by
$H_C = E_C (2N_C + n_d - n_g)^2$, where $n_d = d^\dagger d$ and $n_g$ is a constant controlled by the gate voltage $V_g$.
In contrast to Ref. \onlinecite{hutzen2012} we do not consider any Josephson coupling to a further superconductor.

In this work, we tune $n_g$ to the value $n_g = 2n - \frac{1}{2}$, with $n$ an integer. This results in a charging
ground state degeneracy between the states $(N_C=n,n_d=0)$ and $(N_C=n-1,n_d=1)$, with the next excited states
at $(N_C=n,n_d=1)$ and $(N_C=n-1,n_d=0)$ having an excitation energy $2E_C$, as shown in Fig. \ref{elandscape}. Note that we have neglected any Majorana interaction energy, $H_{int}=\epsilon_m\left(n_d-1/2\right)$, which would break the ground state degeneracy. The energy $\epsilon_m$ is proportional to the Majorana wave function overlap and can be made exponentially small by a sufficiently large system size. Although this must be balanced by the requirement of maintaining a large $E_C$, this is not an issue since degeneracy can be restored by retuning $n_g$ to $n_g=2n-\frac{1}{2}-\frac{\epsilon_m}{2E_C}$. While this retuning does cause a splitting between the first excited states of $4\epsilon_m$, the degeneracy of these states is inessential to our results and as long as $\epsilon_m \ll E_C$ this perturbation to the excited state energy is of no consequence. 

Further excited states appear only at energy $4E_C$ above the first excited states and are neglected
in the present low energy description.
The resulting situation is reminiscent of the large interaction limit of the Anderson model with
two-fold degenerate ground and first excited states, yet with the restriction that there is no
direct scattering process connecting the two excited states because they have different total particle numbers,
$2n-2$ and $2n+1$. This excludes the virtual spin-flip type processes, that dominate Kondo physics, arising
from the usual mapping of the Anderson model on the Kondo model, and the resulting physics for the present
situation is fundamentally different. To discriminate it from the Kondo type behaviour obtained by a mutual coupling
of several such Majorana wires through a common superconductor,
\cite{fu2010,zazunov2011,hutzen2012,zazunov2012,wang2013,ulrich2015,sau2015,plugge2015,plugge2016} and from the behaviour of independent Majorana states,
we call the effective
model obtained from an analogous mapping the \emph{Kondorana model} because it combines Kondo and Majorana
properties on an equal footing, but exhibits exciting new physics. We note in passing that the other charging degeneracy point, $n_g = 2n+\frac{1}{2}$, also results in a many-body state similar to the one found in this work, due to particle-like symmetry.\cite{plugge2015}

We construct the effective model through a Schrieffer-Wolff transformation, whose details are provided
in Appendix A.
As indicated in Fig. \ref{elandscape},
the normal particle tunnelling between leads and Majoranas generates the high energy excitations. Since there is no
direct transition between both excited states, the virtual excitations into the high energy sector generate
an $n_d$ dependent scattering potential between electrons, including a teleportation type scattering
across the Majorana wire $\sim c_{1k}^\dagger c_{2k'}$, \cite{fu2010} but do not cause any change in the Majorana parity.
The resulting effective Hamiltonian is
\begin{align}
	H_\text{eff}
	&= H_{el}
	+  \frac{1}{\sqrt{2}} \sum_k \left[\bigl(J^{(1)}_\pm c_{1k}^\dagger  + i J^{(2)}_\pm c_{2k}^\dagger \bigr)S^+ + \text{H.c.}\right]
\nonumber\\
	&+ \sum_{k,k'}
	\Bigl[
		J_z^{(11)}c^\dagger_{1k}c_{1k'} +J_z^{(22)}c^\dagger_{2k}c_{2k'}
\nonumber \\
	&+	J_z^{(12)} i \bigl(c^\dagger_{1k}c_{2k'} -c^\dagger_{2k'}c_{1k}\bigr)
	\Bigr]S^z,
\label{eq:Heff}
\end{align}
where $S^+ = \sqrt{2} d^\dagger \e^{-i\chi}$, $S^- = \sqrt{2} d \e^{i\chi}$, $S^z = 2n_d -1$
are pseudo-spin operators, and $J^{(j)}_\pm = t_{j}$, $J^{(jj')}_z = t_j t_{j'}/2E_C$ for $j,j'=1,2$. A Zeeman-like term arising from the Schrieffer-Wolff transformation is omitted in Eq. (1) since it can be eliminated in the same way as $\epsilon_m$ (see above and Appendix A).

This Hamiltonian differs from the Kondo Hamiltonian in two essential ways.
Firstly, it cannot be written down as a pure spin-spin interaction because it involves
the tunnelling terms $J_\pm^{(j)}$ which create and annihilate electrons while flipping $S$. Secondly, the $S^z$ term couples in parallel to an $s^z$ and $s^y$ electron
pseudo-spin: Since the tunnelling electrons have a spin polarization locked to the lead,
we can define a lead-spin pseudo-spin with projections
$s \in \{s_+,s_-\} = \{(j=1,\dw),(j=2,\up)\}$
and operators
$s^{\alpha}_{k,k'} = c^\dagger_{ks} \sigma^\alpha_{s,s'} c_{k's'}$ for $\sigma^\alpha$
the Pauli matrices (with $\sigma^0$ the unit matrix).
This allows us to write the $S^z$ term as
$ \sum_{k,k'} [\frac{1}{2} (J_z^{(11)} + J_z^{(22)}) s^0_{kk'} + \frac{1}{2}(J_z^{(11)} - J_z^{(22)}) s^z_{kk'}
+J_z^{(12)} s^y_{kk'}]S^z$. This special form, mainly the appearance of the $s^y_{kk'}$ term,
leads to a behaviour of the Kondorana model that is entirely different from the usual Kondo physics.


\section{Renormalization}  The non-Kondo behaviour of the model becomes evident if we
consider a renormalization group analysis. Since $H_\text{eff}$ describes free electrons
that are coupled to a single localized pseudo-spin $S$, the poor man's scaling technique \cite{anderson1970}
provides a transparent approach to the physics while being accurate.
The renormalization incorporates the modification of the coupling constants $J$ by virtual
scattering processes to high energies. The corresponding diagrams for the $J_\pm^{(j)}$
coefficients are shown in Fig. \ref{jpmscaling}, and details of the calculation are given in Appendix B. We obtain the scaling equations
\begin{equation} \label{eq:scaling}
	\frac{d}{d\ell} \begin{pmatrix} J^{(1)}_\pm \\ J^{(2)}_\pm \end{pmatrix}
	=
	-2 \rho
	\begin{pmatrix}
		 J^{(11)}_z & -J^{(12)}_z \\
		-J^{(12)}_z &  J^{(22)}_z
	\end{pmatrix}
	\begin{pmatrix} J^{(1)}_\pm \\ J^{(2)}_\pm \end{pmatrix},
\end{equation}
where $\ell \sim -\ln(D/D_0)$ and $\rho \sim 1/D_0$, with $D$ the running cutoff energy and $D_0$ the initial electron bandwidth.
From a similar analysis we find that $d J^{(jj')}_z/d\ell = 0$ for all $j,j'$,
which is a consequence of the fact that there are no $S^\pm s^\mp$ terms in $H_\text{eff}$ and there are thus
no $J^{(j)}_\pm$ terms contributing to the $J_z^{(jj')}$ renormalization.

\begin{figure}
	\centering
	\includegraphics[width=\columnwidth]{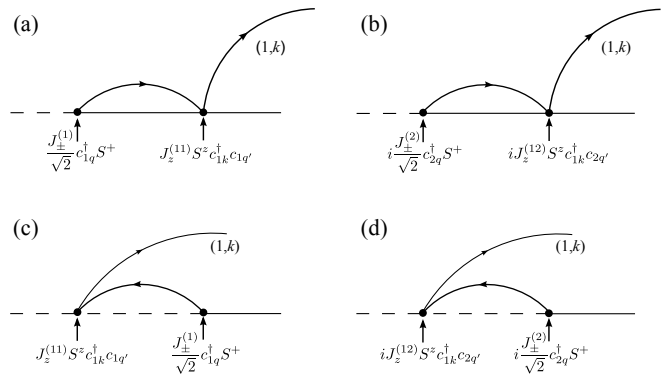}
	\caption{\label{jpmscaling}
		Scattering channels which contribute to renormalization of $J_\pm^{(1)}$. Diagrams (a) and (b) show particle mediated scattering via the left and right leads, respectively. Similarly, (c) and (d) depict hole mediated scattering. Curved lines represent lead electrons, whilst the straight lines correspond to the nanowire with dashed and solid lines denoting $S_z=-1$ and $S_z=+1$, respectively.
	}
\end{figure}

\begin{figure}[h]
	\centering
	\includegraphics[width=\columnwidth]{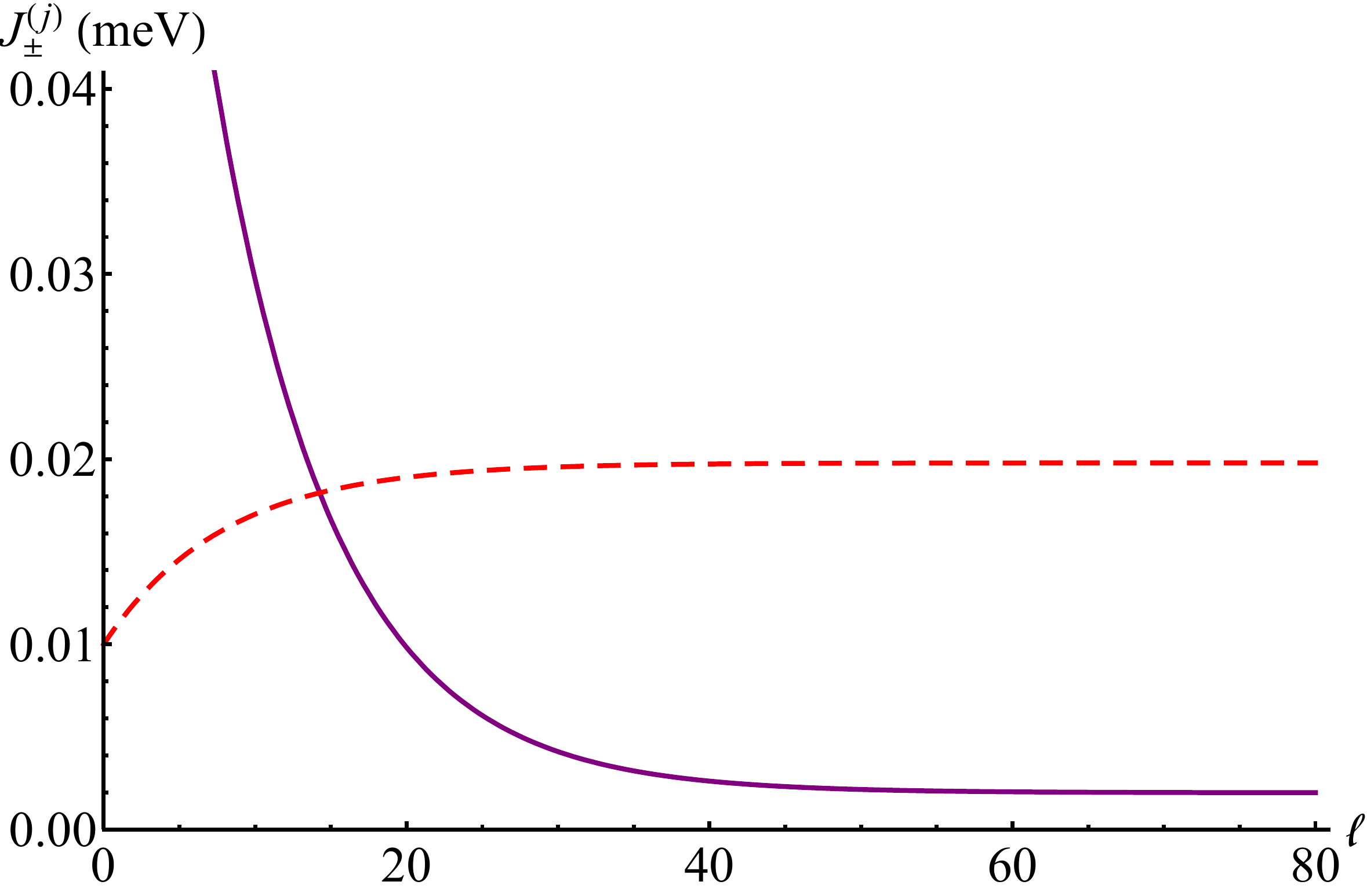}
	\caption{\label{rgflow}
		 Change in anomalous tunnel couplings, $J_\pm^{(j)}$, with scaling parameter $\ell$, as given by Eq. \eqref{eq:scaling_result} for $t_1 = 0.1$meV, $t_2=0.01$meV and further parameters \cite{parameters} based on a recent experiment. \cite{albrecht2016} The solid purple and dashed red lines show $J_\pm^{(1)}$ and $J_\pm^{(2)}$, respectively. The vertical dashed black line is the value of $\ell$ corresponding to the crossover temperature $T_c$. The couplings display a rapid, exponential change from their initial values of $t_1$ and $t_2$ as $\ell$ increases. This high temperature sensitivity even far above $T_c$ is due to the large ratio $t_1/t_2=10$. }
\end{figure}

The renormalization flow of $J^{(j)}_\pm$ is governed by the eigenvalues of the
matrix in Eq. \eqref{eq:scaling}, which remain constant due to the invariance of the $J^{(jj')}_z$.
Since $J^{(jj')}_z = t_j t_j'/2E_C$ we find that the matrix has the eigenvalues $0$ and
$\lambda = J^{(11)}_z + J^{(22)}_z = (t_1^2+t_2^2)/2 E_C$, such that
\begin{equation} \label{eq:scaling_result}
	\begin{pmatrix} J^{(1)}_\pm(\ell) \\ J^{(2)}_\pm(\ell) \end{pmatrix}
	=
	\frac{t_1^2-t_2^2}{t_1^2+t_2^2} \begin{pmatrix} t_1 \\ -t_2 \end{pmatrix} \e^{-2 \rho \lambda \ell}
	+
	\frac{2 t_1 t_2}{t_1^2+t_2^2} \begin{pmatrix} t_2 \\ t_1 \end{pmatrix}.
\end{equation}
The scaling therefore interpolates between the bare $J^{(j)}_\pm$ values and the fixed points
$\bar{J}^{(1)}_\pm = 2 t_1 t_2^2 / (t_1^2 + t_2^2)$ and
$\bar{J}^{(2)}_\pm = 2 t_1^2 t_2 / (t_1^2 + t_2^2)$, as shown in Fig. \ref{rgflow}, and does not display the weak or strong coupling
behaviour of a regular Kondo system. Although the fixed point is finite and the Hamiltonian maintains its form, the resulting state has an involved non-local many-body structure. This is exemplified by the fact that the tunnel coupling asymmetry, $t_1>t_2$, is reversed such that $t_1^*<t_2^*$ at the fixed point, showing that even for local coupling, the entire system including the leads is involved. Indeed, the state revealed above is highly non-local, extending over both leads regardless of nanowire length, and is comprised of lead electrons, Majorana modes and the superconducting condensate. We believe that such a state surpasses the threshold of being merely described as dressed and requires the many-body epithet. 

It must be stressed that the existence of the 0 eigenvalue is a direct consequence of the
$S^z s^y$ term in the Hamiltonian which incorporates the teleportation contribution unique
to the Majorana system, and that the scaling equations \eqref{eq:scaling_result} would be hard
to obtain in any other system. In the absence of the $S^z s^y$ term, the renormalization would flow to the regular weak coupling limit of the ferromagnetic Kondo model.

A significant consideration is whether or not the fixed point $\bar{J}^{(j)}_\pm$ will be reached in practice.
The final value for $\ell$ is determined by the cutoff scale $D$ and the renormalization stops
when $D$ becomes equal to the thermal energy $k_B T$ or any voltage bias applied to the system.
The crossover scale from the bare to the renormalized values is obtained by setting $2 \rho \lambda \ell \sim 1$,
which resolves to
\begin{equation}\label{T_c}
	k_B T_c \sim D_0 \e^{- 1/2\rho\lambda} = D_0 \e^{-E_C D_0/ (t_1^2+t_2^2)}.
\end{equation}
But, since the $J^{(j)}_\pm$ only renormalize for $t_1 \neq t_2$,
this only makes sense for substantially different $t_1$ and $t_2$, as otherwise the changes
in $J^{(j)}_\pm$ are small. Substituting realistic system parameters \cite{parameters} into Eq. \eqref{T_c}, we notice that the flow is very slow, and practically the fixed point is never reached. Due to this it is also of little relevance if the fixed point remains finite when further corrections beyond poor man's scaling are taken into account. Such corrections have an even slower renormalization flow and are always cut off before becoming important.


\section{Transport} A straightforward verification of the behaviour predicted by Eq. \eqref{eq:scaling_result} can be achieved
by measuring the two terminal conductance of the topological superconductor through the Majorana states. Neglecting terms
in Eq.~\eqref{eq:Heff} proportional to $J_z^{(jj')}$, which are a factor $t/E_C$ smaller than the anomalous tunnelling
processes, the effective tunnelling Hamiltonian is given by
$H_T = \sum_k [J_\pm^{(1)}c^\dagger_{1k}f+J_\pm^{(2)}c_{2k}^\dagger f + \text{h.c.}]$, where we define
the composite fermion, $f=d^\dagger e^{-i\chi}$, and where the amplitudes $J_\pm^{(j)}$ are the results of
the renormalization flow.
If we further treat the leads in the wide-band limit, then the situation is exactly analogous to a non-interacting resonant
level model, for which the (peak maximum) differential conductance at $eV<k_BT$ is \cite{haug1996}
\begin{equation}\label{conductance}
	G =K\frac{|J^{(1)}_\pm|^2|J^{(2)}_\pm|^2}{|J^{(1)}_\pm|^2+|J^{(2)}_\pm|^2},
\end{equation}
where $K= \frac{\pi^2 e^2\rho}{h k_BT}$, with $e$ being the electronic charge and $h$ being Planck's Constant. In principle, this conductance offers two signatures of the many-body state found above. First, at constant temperature, $T$, the variation of conductance with changing $t_1,t_2$ asymmetry is markedly different for $T\gg T_c$ and $T\ll T_c$. In the former case, the conductance is $G = K\frac{t_1^2t_2^2}{t_1^2+t_2^2}$,
whereas, at low temperatures, we find that
\begin{equation}\label{lowG}
	G \simeq K\frac{4t_1^4t_2^4\left[t_1^2t_2^2+\left(t_1^2-t_2^2\right)^2\e^{-\alpha}\right]}{\left(t_1^2+t_2^2\right)^3} \hspace{5pt} \text{at $T\ll T_c$},
\end{equation}

\noindent where $\alpha = \frac{\ln(k_BT/D_0)}{\ln(k_BT_c/D_0)}$. However, for realistic system parameters, Eq. \eqref{lowG} implies that, even though $T_c$ may be just about realisable in experiments, the temperature at which true fixed point behaviour is achieved is several orders of magnitude lower.

\begin{figure}[h]
	\centering
	\includegraphics[width=\columnwidth]{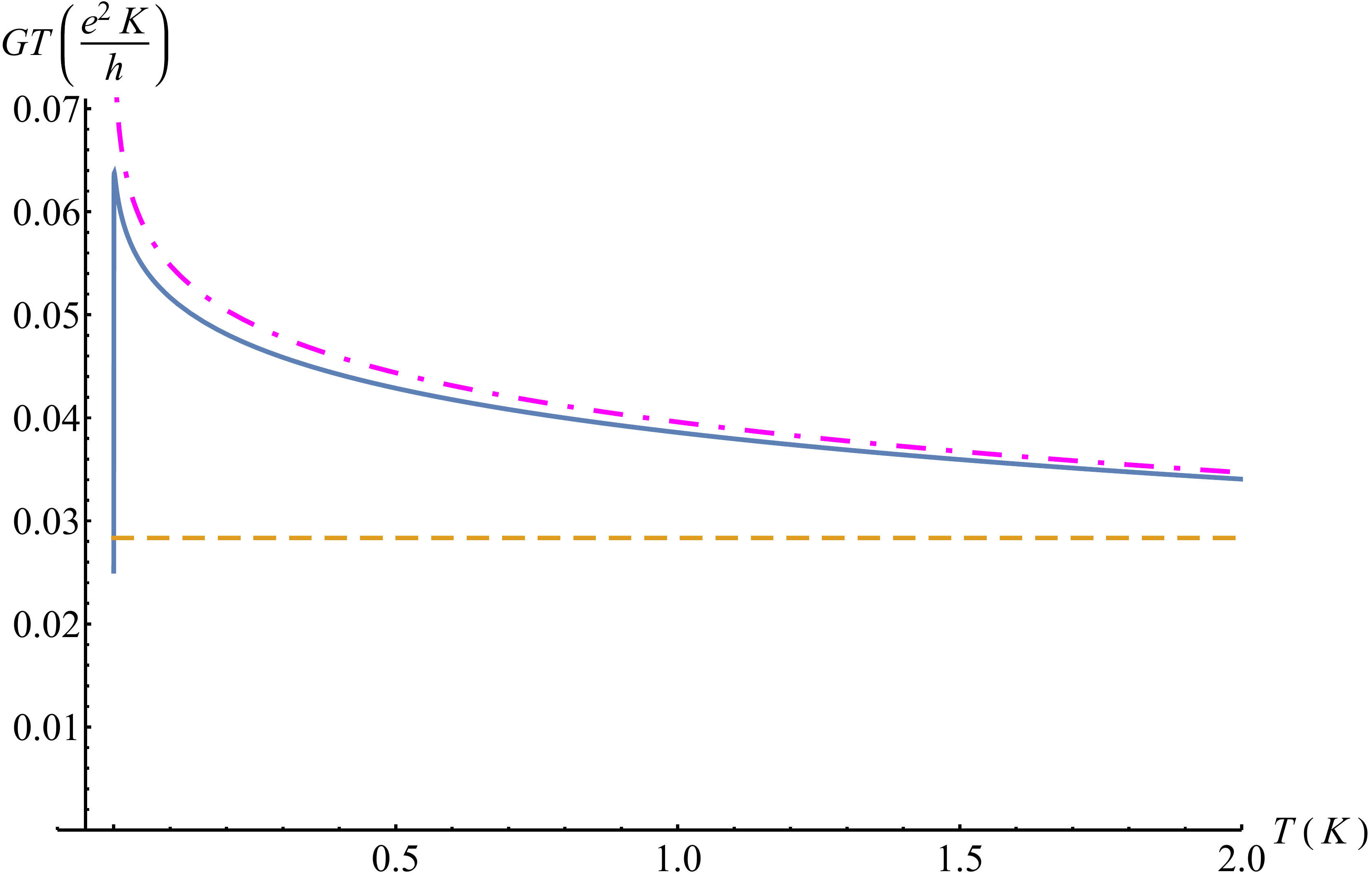}
	\caption{\label{GTvsT}
		Variation of conductance amplitude with temperature, adjusted to account for generic $1/T$ dependence, as given by Eqs. \eqref{eq:scaling_result} and \eqref{conductance}.
		The solid blue line depicts the result for a many-body state whilst the dashed orange line corresponds
		to the bare tunnel couplings. The dot-dashed magenta line is a high temperature ($T\gg T_c$), high asymmetry ($t_1\gg t_2$) approximation, $G=K t_2^2\left(\e^{-2\alpha}-4\e^{-\alpha}+4\right)$, where $\alpha = \frac{\ln(k_BT/D_0)}{\ln(k_BT_c/D_0)}$.  The parameters used \cite{parameters} imply
		$T_c = 2$ mK.
	}
\end{figure}

 We therefore propose a further test for the existence of a many-body state, at $T>T_c$. Fixing the system parameters, but varying $T$, results in a distinctive signature, as shown in Fig. \ref{GTvsT}. Here we plot the product of $G$ and $T$, to remove the direct $1/T$ dependence in Eq. \eqref{conductance}.

It is remarkable that, even at temperatures well above $T_c$, there is a clear difference between the scaled result and the result found from the bare tunnel couplings. That the influence of the many-body state extends to such high temperatures is a result of the strong $J_\pm^{(jj')}$ dependence in both the numerator and denominator of Eq. (6).  Observation of the characteristic behaviour shown in Fig. \ref{GTvsT} appears to be within reach of current experiments and would provide compelling testament to the importance of many-body effects in this system.


\section{Conclusions} 

A floating topological superconductor is an obvious Majorana-based system in which to explore Kondo physics. The Majorana modes of the superconducting condensate are analogous to magnetic impurities or quantum dots in the traditional Kondo effect, whilst the metallic leads fill the role of the electronic continuum.

One might initially expect this system to exhibit something close to conventional Kondo physics, possibly with a slight modification due to the Majorana modes. However, as we have shown in this work, this is not the case. Using a Schrieffer-Wolff transformation and a scaling analysis to eliminate the high energy sectors in the wire and leads respectively, we have demonstrated that the system flows to an intermediate fixed point, rather than the strong or weak coupling of the Kondo model. Indeed, we have shown that the presence of Majoranas is essential for this intermediate scaling to exist. 

The distinct behaviour arising from the interplay of Kondo and Majorana physics motivates our use of the term \textit{Kondorana} to describe our model and the many-body state that arises from it. We have suggested possible experimental transport signatures of this state which are, in principle, within reach of current experiments.


\section{Acknowledgements} We thank C. J. F. Carroll, R. Egger, K. Grove-Rasmussen, C. A. Hooley, M. Scheurer, P. Wahl and A. A. Zyuzin for valuable discussions and comments. IJvB acknowledges studentship funding from EPSRC under grant no. EP/I007002/1.

\section{Appendix A: Derivation of Effective Hamiltonian}

To find an effective low-energy theory of the full Hamiltonian, we carry out a Schrieffer-Wolff transformation
defined by the unitary transformation $H_{\text{eff}} = e^{W}He^{-W}$. This transformation is chosen such that
it eliminates the tunnelling processes into the high energy sector of the model and replaces them by
effective low-energy processes created by virtual high energy excursions. With our choice of tuning the gate
to the degenerate ground states $(N_C=n,n_d=0)$ and $(N_C=n-1,n_d=1)$, the direct tunnelling terms provide the
excitations to the high energy sector, given by the Hamiltonian
\begin{equation}
H_1 = \sum_k ( t_1 c_{1k}^\dagger d + i t_2 c_{2k}^\dagger d ) + \mathrm{H.c.},
\end{equation}
whereas the low energy sector is described by
\begin{align}
H_0 = H_{el} + H_C + \sum_k ( t_1 c_{1k}^\dagger \e^{-i\chi} d^\dagger + i t_2 c_{2k}^\dagger \e^{-i \chi} d^\dagger) + \mathrm{H.c.}
\end{align}
Expanding the unitary transformation in $W$ leads then to the effective Hamiltonian
\begin{equation}
H_{\text{eff}} = H_0 + \frac{1}{2}\left[W,H_1\right],
\end{equation}
in which we have required $\left[W,H_0\right] = -H_1$ such that the first order high energy excitations
vanish. This requirement has the solution
\begin{equation}
W = \sum_k \Xi(\epsilon_k) \bigl(t_1 c^\dagger_{1k}d -i t_2 c^\dagger_{2k}d \bigr) - \text{H.c.},
\end{equation}
with $\Xi(\epsilon_k) = [\epsilon_k-E_C(4N_C+1-2n_g)]^{-1}$.
In deriving this result we have neglected
Andreev type processes of the form $c^\dagger c^\dagger \e^{-i\chi}$. Such processes are indeed generated
at second order in tunnelling, but since they change the number of charges on the wire by 2, they
always lead to high energy excitations and contribute to the low energy theory only at order
$\mathcal{O}(t_j^3/E_C^2)$.

Since is always possible to choose real $t_1, t_2$ (any phase can be absorbed by shifting the
phases of the lead electrons through a gauge transformation), we then find that the
effective Hamiltonian is given by
\begin{align}
&H_{\text{eff}}
=
E_C (2N_C +n_d -n_g)^2
\nonumber\\
&+
\sum_k
\Bigl[
\epsilon_k
\bigl( c^\dagger_{1k}c_{1k}+c^\dagger_{2k}c_{2k}\bigr)
\nonumber\\
&+
\bigl(
t_{1} c^\dagger_{1k}d^\dagger e^{-i\chi}
+
it_{2}c^\dagger_{2k}d^\dagger e^{-i\chi}
+ \text{H.c.}
\bigr)
\Bigr]
\nonumber\\
&+\sum_{k,k'} \Xi(\epsilon_k)
\Bigl[
t_1^2c^\dagger_{1k}c_{1k'}
+
t^2_2c^\dagger_{2k}c_{2k'}
\nonumber\\
&-
\delta_{k,k'}\bigl(t_1^2+t_2^2\bigr) n_d
+
it_1t_2
\bigl(c^\dagger_{1k}c_{2k'}-c^\dagger_{2k}c_{1k'}\bigr)
\Bigr].
\end{align}
The term with $\delta_{k,k'}$ in the last line produces an energy shift for the
$n_d$ level, similar to an overlap integral between the two Majorana wave functions, or a Zeeman-type term for the pseudo-spin $S^z$ in Eq. \eqref{eq:Heff}.
If $\rho(\epsilon)$ is the density of states and $D_0$ is the electron bandwidth
such that $\rho \sim 1/D_0$, we can estimate the magnitude of this term
as
\begin{align}
&(t_1^2+t_2^2)\sum_{k,k'} \delta_{k,k'} \Xi(\epsilon_k)
=
(t_1^2+t_2^2) \int d\epsilon \rho(\epsilon) \Xi(\epsilon)
\nonumber\\
&\sim
-\frac{t_1^2+t_2^2}{D_0} \ln\left[\frac{D_0-E_C(4N_C+1-2n_g)}{D_0+E_C(4N_C+1-2n_g)}\right].
\label{eq:nd_energy_ren}
\end{align}
For $E_C < D_0$ this term is on the order of $\mathcal{O} \left(t_j^2 E_C/ D_0^2\right)$ and thus smaller
than all other considered energies. Yet it can always be removed by a slight adjustment
of $n_g$ through the gate voltage since it plays the same role as the charging energy,
and we shall set it to zero henceforth.

For the remaining effective theory the $\epsilon_k$ term in $\Xi(\epsilon_k)$ is
unimportant since it causes only small corrections for the low-energy properties, and we shall
drop it in the following and use the approximation $\Xi(\epsilon_k) = \Xi = - [E_C(4N_C+1-n_g)]^{-1}$.
We finally notice that with $n_g = 2n+1/2$,
\begin{equation} \label{eq:NC_nd_id}
4N_C+1-2n_g =
\begin{cases}
+2 & \text{for $(N_c=n,n_d=0)$},\\
-2 & \text{for $(N_c=n-1,n_d=1)$},
\end{cases}
\end{equation}
which allows us to write $\Xi = \left(2n_d-1\right)/2E_C$ for these two states.
With these results we find that the effective Hamiltonian
takes the form of Eq. \eqref{eq:Heff}, with the latter identities leading to the
$S^z$ pseudo-spin term. Deviations from Eq. \eqref{eq:NC_nd_id}, such as by tuning $n_g$
slightly away from $2n+1/2$ due to compensation of Eq. \eqref{eq:nd_energy_ren} or due to
the neglected dependence of $\Xi$ on $\epsilon_k$ cause only corrections that either
remain proportional to $S^z$ or are independent of $S^z$ and consist only of renormalizations
of the chemical potentials in the leads. Equation \eqref{eq:Heff} therefore represents the
generic effective Hamiltonian of the system.

\section{Appendix B: Renormalization of Couplings }

\begin{figure}[h]
	\centering
	\includegraphics[width=\columnwidth]{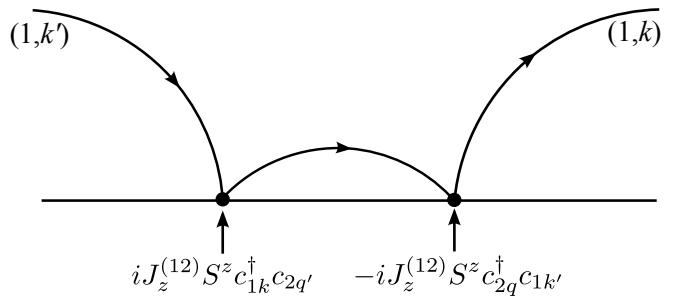}
	\caption{\label{jzscaling} The lowest order particle mediated process contributing to the scaling of $J_z^{(11)}$. The line between the two vertices denotes a particle excited to the high energy shell. Note that the two scattering events commute, so the pathway shown here is exactly cancelled by a corresponding hole-mediated process and does not contribute to scaling.
	}
\end{figure}

Poor man's scaling consists in a renormalization group approach in which excitations to high energy states are successively
integrated out, and the bandwidth is effectively reduced. In the following we label with $q,q'$ these high energy states
and with $k,k'$ the initial and final low energy states. The renormalization proceeds by directly producing corrections
to the Hamiltonian.
We follow a diagrammatic variant of poor man's scaling. The first point to note is that the $J_z^{(jj')}$ couplings are invariant under scaling. The reason for this can be understood by considering Fig. \ref{jzscaling}, which shows the two vertex process contributing to the scaling of $J_z^{(11)}$.

Neither of the two vertices causes a change in Majorana parity, i.e.\ an $S^z$ spin flip, and they therefore commute. The result is that the hole-mediated version of the depicted process will result in exact cancellation. Indeed, since only terms in the Hamiltonian proportional to $J_\pm^{(j)}$ change $S^z$, and such terms constitute terminal vertices, as shown in Fig. \ref{jpmscaling}, there are no scattering diagrams, to any order, that result in scaling of $J_z^{(jj')}$.

We now turn to the scaling of $J_\pm^{(1)}$, for which the first order scattering processes are shown in Fig. \ref{jpmscaling}. The particle mediated channels, with excitations $q,q'$ into an energy shell of width $\delta D$ at the upper band edge, lead to the following correction of the Hamiltonian,
\begin{align}
&\delta H_p = \sum_{q,q'}\Bigl[
J_z^{(11)}S^zc^\dagger_{1k}c_{1q'}\left(E-H_0\right)^{-1}J_\pm^{(1)}c^\dagger_{1q}\frac{1}{\sqrt{2}}S^+
\nonumber\\
&+
iJ_z^{(12)}S^zc^\dagger_{1k}c_{2q'}\left(E-H_0\right)^{-1}iJ_\pm^{(2)}c^\dagger_{2q}\frac{1}{\sqrt{2}}S^+
\Bigr]
\nonumber\\
&= \sum_{q,q'}\Bigl[
J_z^{(11)}J_\pm^{(1)}c^\dagger_{1k}c_{1q'}\left(E-D\right)^{-1}c^\dagger_{1q}\frac{1}{\sqrt{2}}S^+
\nonumber\\
&-
J_z^{(12)}J_\pm^{(2)}c^\dagger_{1k}c_{2q'}\left(E-D\right)^{-1}c^\dagger_{2q}\frac{1}{\sqrt{2}}S^+
\Bigr],
\end{align}
where $E$ is the energy at which the system is probed and $D$ is the running bandwidth of the leads.
We have used the fact that $S^zS^+=S^+$ and written $H_0c^{\dagger}S^+=Dc^{\dagger}S^+$, since $|\delta D| \ll D$ and $S^+$ corresponds to zero energy excitations. Summing over the high energy interval $|\delta D|$, using the fact that $E-D \simeq -D$ and that
far above the Fermi surface $c_{1q'}c^\dagger_{1q} = \delta_{qq'}$, we find that the particle mediated contribution to the scaling is
\begin{equation}
\delta H_p = -\frac{\rho|\delta D|}{D}\left[J_z^{(11)}J_\pm^{(1)}-J_z^{(12)}J_\pm^{(2)}\right]c^\dagger_{1k}\frac{1}{\sqrt{2}}S^+,
\end{equation}
where $\rho$ is the density of states in the leads. A similar analysis for the hole mediated terms provides an identical result,
such that the total Hamiltonian associated with the two vertex events corresponding to $J_\pm^{(1)}$ is therefore
\begin{equation}
\delta H_{\text{2v}}
= -\frac{2\rho|\delta D|}{D}\left[J_z^{(11)}J_\pm^{(1)}-J_z^{(12)}J_\pm^{(2)}\right]\frac{1}{\sqrt{2}}c^\dagger_{1k}S^+.
\end{equation}
Comparing this with Eq. \eqref{eq:Heff}, we see that renormalization group flow equation for $J_\pm^{(1)}$ is
\begin{equation}\label{eqn:1flow}
\frac{d}{d\ell}J_\pm^{(1)} = - 2\rho\left[J_z^{(11)}J_\pm^{(1)}-J_z^{(12)}J_\pm^{(2)}\right].
\end{equation}
The derivation of the scaling for $J_\pm^{(2)}$ is essentially identical and, with Eq. \eqref{eqn:1flow} gives the result shown in Eq. \eqref{eq:scaling}.
\vfill


\end{document}